# Software Carpentry – get more done in less time

Alexandra Simperler (1) and Greg Wilson (2)

(1) Imperial College London/ a.simperler@imperial.ac.uk
(2) Software Carpentry / gvwilson@software-carpentry.org

**Abstract**

The aim of this study was to investigate if participants of Software Carpentry (SC) get more done in less time. We asked 32 questions to assess 24 former participants to analyse if SC gave them the computing skills to accomplish this. Our research shows that time was already saved during the workshop as it could shorten the learning process of new skills. A majority of participants were able to use these new skills straight away and thus could speed up their day to day work.

# 1. Introduction

Software Carpentry (SC) [1,2] was founded in 1998, and is delivered as a two day on-site workshop. The target audience are researchers who are involved in computer assisted science, engineering, medicine, and related disciplines. SC delivers core skills to enable participants to acquire the computing skills they need to get more done in less time, and with less difficulty. A typical workshop comprises the Unix shell (and how to automate repetitive tasks), Python or R (and how to grow a program in a modular, testable way), Git and GitHub (and how to track and share work efficiently) and SQL (and the difference between structured and unstructured data). These topics give a good basis of computer skills, which can be easily expanded. With this study we wanted to investigate if participants indeed get more done in less time. We constructed a catalogue of 32 questions (see Appendix 1) and interviewed 24 former SC participants. We asked questions to learn about the participants themselves and to see who is attracted to SC. We wanted to know how easily they found their way to a SC workshop, and thus, establish if SC is a brand already and an obvious choice for the target group we have in mind. It was important for us to learn about the participants' pre-course expectations, to see if SC caters for the right people or needs adjustment. The most important thing was to analyse if indeed participants get more done in less time, and our study reveals that this is the case. The SC community are growing their own talents – people who are attending courses often get involved as instructors, tutors and organisers.

# 2. About the participants

We send an email to former participants and described the study to them. We then interviewed 24 former participants of SC who responded and were willing to take part. All of the interviewees volunteered without being offered compensation. They were motivated by being offered that their contributions may help to improve SC for future cohorts of participants. The interviews were conducted in person or via telecommunication tools by Alexandra Simperler, who is not employed by SC.

The following tables comprise particulars about the participants of this study:

| Country | Number of participants |
|---|---|
| UK | 20 |
| US | 2 |
| Ireland | 1 |
| Norway | 1 |

| Age/yrs | Number of participants |
|---|---|
| <26 | 3 |
| 26-35 | 15 |
| > 35 | 6 |

| Gender | Number of participants |
|---|---|
| female | 8 |
| male | 16 |

The participants consisted of PhD students, postdoctoral researchers, academics and company employees. Their fields of work were mainly scientific; they were working in the fields of computer assisted life, geological, engineering, physical, and library sciences and programming.

| How long before the interview did they attend SC? | Number of participants |
|---|---|
| < 3 month | 5 |
| 3 – 12 months | 12 |
| > 12 months | 7 |

A third of them attended with a colleague, which means SC appeals to individual participants as well as to peer groups. Often, SC seems to bring together people from the same work organisation who have never met before as they were from different departments.

We were interested in the networking ability of the participants as we think that the process of acquiring contemporary IT skills profits very much from interaction with peers, i.e. collaborative learning [3]. 65% of the people met a new person at the workshop. 11 people (slightly less than 50%) could retain contact via internal work networks or social media.

Overall, gender, age and occupation did not influence the answers given. What influenced the answers most, was how much formal training in computing people had received before:

| Mode of acquiring computer skills | Number of participants |
|---|---|
| Formal training | 2 |
| Self-training | 13 |
| Combination of formal and self-training | 9 |

**3. Software Carpentry- a brand?**

SC was founded in 1998, so it should be a known entity. About 60% of all people did not know about Software Carpentry before enrolling on a workshop. However, 15 people were told by colleagues or their work's internal email lists, 3 learned about it from social media and networking events, and one was engaged as a trainer due to their professional knowledge. Only 5 people were actively looking for workshops, and their internet searches led them to a SC boot camp close to their place of residency.

| I found out about the SC boot camp, because | Number of participants |
|---|---|
| … my colleagues or work place told me | 15 |
| … of social media | 3 |
| … I was asked to help out | 1 |
| … I was searching the internet and it came up | 5 |

When asked what comes into their minds first when they hear "Software Carpentry" and what the workshop meant to them, we found very individual answers. 11 people had "workshop" in mind and, thus, went straight for SC's definition. 4 people had "experience" in mind and named adverbs such as "useful" and "good". 8 people's first thoughts were with respect to SC's philosophy of improving code by giving it improved structure and by being in control. One interviewee was thinking of SC's role as an interface between a researcher, who is not formally trained in coding, and programming itself. SC certainly has developed a good name and can be easily found by people with an intention of improving their skills.

| Software Carpentry is … | Number of participants |
|---|---|
| …a workshop | 11 |
| … useful | 4 |
| … helpful to improve my code | 8 |
| … an interface between scientists not formally trained and programming | 1 |

### 4. Motivation

When analysing the question, why people wanted to attend SC, we recognized three main motives with our interviewees: interest, continuous professional development (CPD), and attending for a specific task. Five people were simply interested in SC as its philosophy aligned with their ideas. These people were keen on networking and heard about SC via that route. Six people saw in SC a good vehicle for their CPD. They talk to colleagues and do react to offerings of training in their workplace's internal mailing lists. Eight were skilled in coding and programming and they wanted to add best practices to their skill set. They specifically looked for courses and got input from their peers where best to obtain these skills. Five people were interested in SC to learn specific skills, for example version control, Python, GitHub, etc. and found these mentioned in SC's workshop outline.

| I was motivated by … | Number of participants |
|---|---|
| … SC's ideas | 5 |
| … CDP in general | 6 |
| … best practises | 8 |
| … specific topics | 5 |

## 5. Learning outcome

A good assessment for learning outcomes is to check what people remember, what new skills they acquired, what skills they improved, and if they considered that the quality of their coding has improved. Our interviewees between them could list and remember all topics SC would usually teach. Git/GitHub and Python were mentioned most often (15 times) followed by the UNIX shell and version control, which was mentioned separately from GitHub. Two people remembered the methods of teaching that were used to deliver the workshop's content.

For 11 people, Git/GitHub was something they never used before. Git [4] (started 2005) is a free and open source distributed version control system and GitHub [5] (started 2008) is a repository for codes. Both are very appealing to people who develop code as a joint venture. Six people specifically mentioned new tips and tricks they were able to pick up on a variety of topics. Version control, unknown UNIX commands, code testing, and Python[6] were mentioned four times each. When we asked people which of their skills they improved, a third mentioned Python. In the last few years in Higher Education it is a language often suggested to people who wish to use programming or scripting for their work [7,8]. Many science students will even find programming courses on their syllabus nowadays. So it is good to see from our analysis that SC can offer an improvement of participant's Python skills. Five people profited from learning more about how to structure code better in general. People who knew about Git, testing, and version control were able to improve their skills as well. Six people managed to learn new shell commands.

More than two thirds of people said that the workshop improved their coding. They managed to write tidier code, to implement some changes, tried to comment better, and reported that they were doing more testing. Three people are planning to improve their coding in the future, which was dependent on the stage of their projects, they are working on. Three people felt that they were up to scratch but that they profited from tips and tricks. Thus, from the given answers we deduced that in all cases Software Carpentry was able to add to or improve the participants' skills. However, we still ask people for their opinion. When asked, if their time was usefully spent on Software Carpentry, 20 people answered that with a definite "yes". Two people felt, the workshop was too advanced for them, so in their opinion they did not improve as much as they liked. Two people felt that their time was mostly useful spent. The workshop was covering topics they knew about already which left them idle at times.

**6. Did they do it in less time?**

To get more done in less time, a workshop should shorten the learning process of new skills as well as enable participants to use new skills straight away. Seven people stated that it would have taken them "forever" to learn these new skills. This was, because they had to find out about the skills first, probably randomly acquire them and also find the motivation and time to do so. For 11 people it would have taken several weeks, whereas six people were confident it would have taken them several days.

| Without SC it would have taken me … | Number of participants |
| --- | --- |
| … "forever" | 7 |
| … several weeks | 11 |
| … several days | 6 |

Thus, in all cases Software Carpentry could shorten and for some participants, even start the learning process.

Instructors and tutors should have a big contribution to enhance and speed up the learning process as well. All but one of the 24 interviewees told us that the SC tutors and instructors helped them to get, where they wanted to be, faster. This one participant attended a workshop with a low number of tutors, i.e. less than one tutor per 10 participants. However, at workshops with a good tutor-participant-ratio there were better one-to-one interactions. Thus, participants were able to profit from discussions, anecdotes, and tips and tricks.

18 people (66%) felt that they got all their questions answered. Two people did not have any questions and four people had some of their questions answered. 8 people (33%) felt a little bit rushed when they tried to recap their learning during the workshop. This may be fixed in the future by correcting the instructors' pace. SC teaches instructor-led, i.e. "live-typing" with the audience following. Novice teachers are perfect for this – they make more mistakes, have to correct them, and have to think if the code they wrote, is actually producing what it should. This gives the participants natural breaks to catch up as well as learn from mistakes. However, every person familiar with teaching software/coding knows that once the courses become familiar, there are less mistakes and a quick glance replaces careful checking what code produces on the command line. Some carefully placed breaks can avoid this and allow participants to digest their learning. However, two of our interviewees took things into their own hands, and came back for a second helping of SC. Another option suggested by participants in the interviews was to split SC into beginners and advanced versions.

All but five people were able to use some new skills soon after attending SC. The reasons stated for not using new skills were that one person would require more freedom in their work environment to implement changes, two people would have to start their project first, and two people need to learn more. Most notably amongst the 19 people reporting quickly using newly acquired skills was: the implementation of version control by two people, one person changed from their preferred programming language to Python, three people started to use GitHub and one person told us that their coding became faster. The remaining 12 people could use some of the little tips and tricks, the instructors shared with them. Thus, 50% of participants could have come out of SC with only little enthusiasm for trying new things soon if it was not for the instructors. It was seen as very positive where instructors could convey how a particular tip helped them with a computational problem. Hence, all SC instructors may be encouraged to keep sharing tips and tricks in that fashion.

| What skills did you use soon after attending | Number of participants |
| --- | --- |
| None (yet) | 5 |
| Version control | 2 |
| Python | 1 |
| GitHub | 3 |
| Faster/more efficient coding | 1 |
| Little tips and tricks | 12 |

So it seems, if our participants are not held back by things out of their control, they are able to do new things, as little or big they may be, soon after attending Software Carpentry.

**7. Mindsets and new ideas**

Participants are usually feel comfortable in a workshop if it agrees with certain practices they are used to. But what a good workshop should do is open people's minds [9,10] to new practices and workflows, and it also should give participants new ideas. Ideally, they should practice these new ideas and the new skills. We also wanted to analyse if they understand and solve more computer-related problems after the workshop which goes together with a change in the mindset.

Many of the participants described themselves as very open and not having too many preconceptions. Some of them are very used to training and are very professional about adapting their ways of thinking. Thus, there was no need for them to change their mindsets, as they could tune in quite easily. One person felt anxious to start with, but the accessibility of instructors/trainers converted their participation into a good experience in the end. People felt, they were given an opportunity to take new things on board. Five people who wished to be more systematic with their programming took away new concepts, structure and best practises[10]. The latter was giving them new ideas and people felt more empowered to improve and control their code. What was quite noticeable was that they started to consider other people or their collaborators who may use the code or even write parts of it. All new ideas participants had, were very individual due to their different work contents. However, what these ideas had in common was to help third parties to understand the individual's work better. There is another important factor of getting things done faster – removing hurdles for collaborators helps immensely to actively drive software usability.

Only three people did not have time yet to practice the learning. This had to do with them being confident enough to immediately implement ideas, or with the fact that SC was attended so recently that they had no time to practice it.

**8. Better self training?**

Another factor to make people faster in acquiring skills is to give them ideas of how they can help themselves, i.e. utilise self-directed learning[12]. Three people felt that they always could solve software problems by themselves. 21 people felt that SC helped them to look at problems in a different way and to feel a bit more confidence in doing so. 10 people actually were fully aware of educational or problem-solving computer resources such as Stack overflow, literature and classes while others thought they felt more confident in using them after attending SC.

**9. Good for others?**

The best way of appreciating a workshop is actually to recommend it to other people or even get involved. All participants indicated that they would have no problem to recommend SC to other people when given the chance, and 23 had done so already. We were asking if our interviewees thought that Software Carpentry should be an essential part in the CPD of professionals who need software skills for their work. People generally liked the idea, but they think it needs to be voluntarily and can be good for undergraduate students. 10 people became involved as trainers and demonstrators, and two of them organised SC's at their work place. 50% of people are or are thinking about of getting involved with the Software Carpentry network.

**10. Conclusion**

Software Carpentry's statement of getting more done in less time is promising. All of our participants saved time when acquiring new skills during the workshop. They also will save time in their future ventures as SC could give them ways to establish or revisit principles behind their scripting/coding. Especially, people who develop software for a community or within a collaboration picked up ideas and tips how to make their code sustainable. A better structure and commenting of a code helps a potential user immensely and prevents the developer from adopting a support role, and thus losing valuable time which may be spend on developing. When working within a project, tracking and sharing work efficiently will help team members to save time, as a team's joint coding effort become more transparent and examinable.

**Acknowledgement**

Alexandra Simperler is grateful to the Software Sustainability Institute for awarding a Fellowship in 2014 to enable this study. We are thanking Mike Jackson from the Edinburgh Parallel Computing Centre (EPCC) for assisting with finding interviewees by forwarding our mail to his mailing lists. We also thank Jory Schossau (Michigan State University), Jason Williams (Cold Spring Harbor Laboratory) and Lee Whitmore (Birkbeck, University of London) for helpful discussions. Finally, we are very grateful to all interviewees who shared their opinions with us and made this study possible.

**References**

[1] "Software Carpentry: Getting Scientists to Write Better Code by Making Them More Productive" G. Wilson; Comput. Sci. Eng. 8 (**2006**) 66. doi:10.1109/MCSE.2006.122

[2] "Software Carpentry: Lessons Learned", G. Wilson, arXiv:1307.5448, July **2013**.

[3] "What do you mean by collaborative learning?" P. Dillenbourg, Collaborative-learning: Cognitive and Computational Approaches., Oxford: Elsevier, pp.1-19,**1999**. <hal-00190240>

[4] https://git-scm.com/

[5] https://github.com/

[6] https://www.python.org/

[7] "Python and Roles of Variables in Introductory Programming: Experiences from Three Educational Institutions." Nikula, U., Sajaniemi, J., Tedre, M. & Wray, S. Journal of Information Technology Education: Research 6 (**2006**) 199-214.

[8] "Python prevails." V. Leping, M. Lepp, M. Niitsoo, E. Tonisson, V. Vene, and A. Villems. Proceedings of the International Conference on Computer Systems and Technologies and Workshop for PhD Students in Computing, CompSysTech '09, pages 87:1–87:5, New York, NY, USA, **2009**. ACM.

[9] "Manipulating Mindset to Positively Influence Introductory Programming Performance." Q. Cutts, E. Cutts, S. Draper, P. O'Donnell, and P. Saffrey, Proceedings of SIGCSE (**2010**) 431-435. doi:10.1145/1734263.1734409

[10] "Dangers of a Fixed Mindset: Implications of Self-theories Research for Computer Science Education." L. Murphy and L. Thomas, Proceedings of ItiCSE (**2008**) 271-275.doi:10.1145/1597849.1384344

[11] "Best Practices for Scientific Computing" G. `Wilson, D.A. Aruliah, C. Titus Brown, N.P. Chue Hong,`
M. Davis, R.T. Guy, S.H.D. Haddock, K.D. Huff, I.M. Mitchell, M.D. Plumbley, B. Waugh, E.P. White, P. Wilson; PLoS Biol 12(1): (**2014**) e1001745. doi:10.1371/journal.pbio.1001745

[12] "Andragogy and Self-Directed Learning: Pillars of Adult Learning Theory." S.B. Merriam, New Directions for Adult and Continuing Education **2001**, p 3–14.

**APPENDIX 1 – Questions to analyse the Software Carpentry (SC)**

1. Your Occupation is …?
2. Did you have formal education in computing/programming /IT or are you self-taught?
3. What is the first thing that comes into mind when you hear SC?
4. Which topics do you remember?
5. Did you feel your time was usefully spent on the SC?
6. What exactly was the SC for you?
7. When did you do the course?
8. Why did you do the course?
9. Did you come with colleagues?
10. How did you find out about this course?
11. Was SC a familiar name?
12. Did you tell other people about SC?
13. Did you get involved with the SC? As trainer, blogger, advocate, …?
14. Do you recommend SC?
15. Could you envisage the SC as a "must have been there"?
16. Did you learn new skills? If yes, which ones?
17. Did you improve skills you had before? If yes, which ones?
18. Did SC encourage you to recap your learning during the event?
19. Could SC change your opinion you had about best practises/workflows/coding/ etc? If yes, …
20. Did the course agree with your mindset/practises/ideas/?
21. Did SC give you new ideas? If yes, …
22. Did SC improve the quality of your coding? If yes, …
23. Was SC teaching you skills you could use the next day, will use in the future?

24. Without the SC how long would you have taken to acquire these skills? Hours, days?

25. Did SC instructors help you to get where you wanted to be faster?

26. Where all your questions answered or did you come out with more than you had when you went in?

27. Were you able to understand and solve more computer-related problems you encountered after SC?

28. Was it easier to use educational or problem solving computer resources after SC? (such as Stack Overflow, books, classes, …)

29. Did you specifically practice or further explore any SC skills after the workshop?

30. Did you meet new people? Job-related

31. Are you still in contact with them? Job-related